\documentclass{JHEP3}
\usepackage{epsfig}

\def\lsi{\raise0.3ex\hbox{$<$\kern-0.75em\raise-1.1ex\hbox{$\sim$}}}
\def\gsi{\raise0.3ex\hbox{$>$\kern-0.75em\raise-1.1ex\hbox{$\sim$}}}

\title{Critical point of 
QCD at finite $T$ and $\mu$, lattice results for physical quark masses}

\author{Z.~Fodor$^{a,b}$,
        S.D.~Katz$^a$\thanks{On leave from Institute for Theoretical
          Physics, E\"otv\"os University, Hungary}\\
    $^a$Department of Physics, University of Wuppertal, Germany\\
    $^b$Institute for Theoretical Physics, E\"otv\"os University, Hungary}

\abstract{ 
A critical point (E) is expected in QCD on the temperature ($T$) versus 
baryonic chemical potential ($\mu$) plane. Using a recently proposed lattice 
method for $\mu\neq$0 we study dynamical QCD with $n_f$=2+1 
staggered quarks of physical masses on $L_t=4$ lattices. Our result 
for the critical point is 
$T_E=162 \pm 2$~MeV and $\mu_E= 360 \pm 40$~MeV. For the critical 
temperature at $\mu=0$ we obtained $T_c=164 \pm 2$ MeV. This work 
extends our previous study [Z. Fodor and S.D.Katz, JHEP 0203 (2002) 014] 
by two means. It decreases the light quark masses ($m_{u,d}$) by a factor 
of three down to their physical values. Furthermore, in order to approach 
the thermodynamical limit we increase our largest volume by a factor of 
three. As expected, decreasing $m_{u,d}$ decreased $\mu_E$. 
Note, that the continuum extrapolation is still missing}


\preprint{ITP-Budapest 609\\ WUB 04-04}

\begin{document}
\section{Introduction.}
 
QCD is an asymptotically free theory, thus its high temperature 
and high density
phases are dominated by partons (quarks and gluons) as degrees of
freedom rather than hadrons. These phases  play an important 
role in the particle physics of the early universe, of neutron stars 
and of heavy ion collisions 
(for a clear theoretical introduction and a review see 
\cite{Wilczek:1999ym,Rajagopal:2000wf}).

Extensive experimental work has been done
with heavy ion collisions at GSI, CERN and Brookhaven 
to study the strong interactions at high temperatures and non-vanishing 
baryon densities (the latter correspond to non-vanishing baryonic
chemical potentials) and to explore
the $\mu$-$T$ phase diagram. 
It is a long-standing open question, whether a critical point (E)
exists on the $\mu$-$T$ plane, 
and particularly how to predict theoretically its location 
\cite{Halasz:1998qr,Berges:1998rc}. 
At this point E the phase transition is of
second order and long wavelength fluctuations appear, which 
results in characteristic experimental consequences, similar to
critical opalescence. Passing close enough to ($\mu_E$,$T_E$)  
one expects simultaneous appearance of 
signatures. The observables exhibit non-monotonic dependence on the 
control parameters \cite{Stephanov:1998dy}, 
since one can miss the critical point on either of two sides.
 
The location of this critical point is 
an unambiguous, non-perturbative prediction of the QCD Lagrangian. 
Unfortunately, until recently no
{\it ab initio}, lattice analysis based on QCD was done to locate
the endpoint. Models 
with infinitely large strange quark mass ($m_s$=$\infty$) 
were used (e.g. \cite{Halasz:1998qr}), 
suggesting that $\mu_E \approx$ 700~MeV.
The result is sensitive to the strange quark mass 
($\mu_E$ should be smaller for smaller $m_s$).
For realistic cases these 
techniques can not predict the value of $\mu_E$ 
even to within a factor of 2-3. 

Lattice QCD at non-vanishing baryon density should, in principle,
give an unambiguous answer. Though QCD at finite $\mu$ can be
formulated on the lattice \cite{Hasenfratz:1983ba}, standard
Monte-Carlo techniques can not be used at $\mu \neq 0$. The reason
is that for non-vanishing real $\mu$ the functional measure
--thus, the determinant of
the Euclidean Dirac operator-- is complex. This fact
spoils any Monte-Carlo technique based on importance sampling.
Several suggestions were studied earlier to solve the problem.
Unfortunately, none of them was able to give the phase line
or locate ($\mu_E$,$T_E$). About two years ago new techniques appeared,
with which moderate chemical potentials could be reached
on the lattice.

In two recent papers we proposed a new 
method \cite{Fodor:2001au,Fodor:2001pe} to study
lattice QCD at finite $T$ and $\mu$. The 
idea was to produce an ensemble of QCD configurations at
$\mu$=0 and at the corresponding transition temperature 
$T_c$ (or at any other physically motivated
point for which importance sampling works). Then we determined
the Boltzmann weights \cite{Ferrenberg:1988yz} 
of these configurations at $\mu\neq 0$
and at $T$ lowered to the transition temperatures at this
non-vanishing $\mu$. An ensemble of configurations at a transition  
point was reweighted to an ensemble of configurations at
another transition point. With this technique a much better
overlap was observed than by reweighting pure hadronic ensemble     
to a transition one \cite{Barbour:1998ej}. 
We illustrated the applicability 
of the method in $n_f$=4 dynamical QCD \cite{Fodor:2001au}
and in $n_f$=2+1 dynamical QCD
\cite{Fodor:2001pe}. The phase line with
the critical end-point \cite{Fodor:2001pe} and the
equation of state \cite{Fodor:2002km,Csikor:2004ik} were determined. 

A less CPU demanding, truncated version of the 
overlap-improving multi-parameter reweighting
was also applied using an improved lattice action. The approach
can be summarized as follows.
Instead of evaluating the Boltzmann weights exactly one can
expand it in the chemical potential and use the first terms
of the expansion. This modified technique was also able
to give the phase diagram \cite{Allton:2002zi} and the
equation of state \cite{Allton:2003vx} (estimates based on
derivative extrapolation for the chiral endpoint was also 
reported in conference proceedings
\cite{Karsch:2003va,Ejiri:2003dc}). The success of the 
overlap-improving multi-parameter reweighting was 
analyzed in Ref. \cite{Ejiri:2004yw}. 

A completely independent method is based on the the fact
that at imaginary chemical potential importance sampling works.
The result on the phase digram can be analytically continued
to real chemical potentials. This technique gave the phase
line both in $n_f$=2 \cite{deForcrand:2002ci}
and in $n_f$=4 \cite{D'Elia:2002gd}, which are consistent
with the results of the overlap-improving multi-parameter 
reweighting method \cite{Fodor:2001au,Fodor:2001pe,Allton:2002zi}.
Analytic continuation was also used to estimate the location of the 
critical endpoint \cite{deForcrand:2003hx} for $n_f$=3.
  
Recently several other new techniques were suggested (see e.g. 
\cite{Hong:2003zq,Liu:2003wy,Ambjorn:2002pz}), which will be
most probably tested in the near future. 
The recent developments of lattice QCD at non-vanishing chemical 
potentials are reviewed by Refs. 
\cite{Kogut:2002kk}.

In this paper we use our original suggestion 
\cite{Fodor:2001au,Fodor:2001pe} and evaluate the reweighting
Boltzmann factors exactly. We determine
the volume (V) dependence of the zeros of the partition function 
on the complex gauge coupling ($\beta$) plane. Based on this volume
dependence we determine the type of the transition as a function of
$\mu$. The endpoint $\mu_E$ is given by the
value at which the crossover disappears and finite volume
scaling predicts a first order phase transition. These finite
$T$ calculations are done on $L_t=4$ lattices. In order 
to set the physical scale we determine the pion, kaon and rho masses
($m_\pi,m_K,m_\rho$), and the
Sommer \cite{Sommer:1994ce} scale ($R_0$) at $T$=0. 
Our quark masses are realistic, the strange quark mass and 
the light quark masses are
set about to their physical values.
Having determined the lattice spacing we transform
our result to physical units and give $T_c$,
the location of ($\mu_E$,$T_E$) and show the phase diagram separating
the hadronic phase and the QGP.

The present work is a significant improvement
on our previous analysis by two means. We increased the physical volume
by a factor of three and decreased the light quark masses  
by a factor of three, down to their physical values. Due to
these improvements the computational effort of the present work
was 140 times larger than that of the previous analysis 
\cite{Fodor:2001pe}. As expected, decreasing the light
quark masses resulted in a smaller $\mu_E$. Increasing the
volumes did not influence the results, which indicates
the reliability of the finite volume analysis. In order to give  
the final answer to ($\mu_E$,$T_E$) the most important
step remained is the continuum extrapolation. Note, however, that 
for the present physical problem decreasing the lattice spacing by a 
factor of 2 increases the CPU costs by approximately three orders
of magnitude.

The remaining part of the paper is organized as follows. In Section 2 we
summarize the overlap-improving multi-parameter reweighting
technique and the method of the Lee-Yang zeros, which can be used 
to separate the crossover and first order transition regions.
Section 3 contains the details of the T$\neq$0 and T=0 simulations.
Those who are not interested in the lattice details could 
directly move to Section 4, in which we present our results and conclude.

\section{Overlap-improving multi-parameter reweighting at $\mu\neq 0$.}

The partition function of lattice QCD with $n_f$ degenerate 
staggered quarks (for an introduction see e.g. \cite{Montvay:cy}) is
given by the functional integral of the bosonic action
$S_{b}$ at gauge coupling $\beta$ over the link variables $U$, 
weighted by the determinant of the quark matrix $M$, which can be 
rewritten \cite{Fodor:2001au} as
\begin{eqnarray} \label{reweight}
&&Z(\beta,m,\mu)=
\int{\cal D}U \exp[-S_{b}(\beta,U)][\det M(m,\mu,U)]^{n_f/4}
\nonumber\\
&&=\int {\cal D}U \exp[-S_{b}(\beta_w,U)][\det M(m_w,\mu_w,U)]^{n_f/4}
\\
&&\left\{\exp[-S_{b}(\beta,U)+S_{b}(\beta_w,U)]
\left[{\det M(m,\mu,U)  \over \det M(m_w,\mu_w,U)}\right]^{n_f/4}\right\},
\nonumber
\end{eqnarray}
where $m$ is the quark mass, $\mu$ is the quark chemical potential 
and $n_f$ is the number of flavours. 
For non-degenerate masses one uses simply 
the product of several quark matrix
determinants on the $1/4$-th power. Standard
importance sampling works and can be used to collect an 
ensemble of configurations at  $m_w$, $\beta_w$ and $\mu_w$ (with  
e.g. Re($\mu_w$)=0 or non-vanishing isospin chemical potential). 
It means we treat the terms in the curly 
bracket as an observable
--which is measured on each independent configuration--
and the rest as the measure. By simultaneously changing 
several parameters e.g.
$\beta$ and $\mu$ one can ensure that even the mismatched measure
at $\beta_w$ and $\mu_w$ samples 
the regions where the original integrand with $\beta$ and $\mu$ 
is large. In practice the determinant is evaluated at some $\mu$ and 
a Ferrenberg-Swendsen reweighting \cite{Ferrenberg:1988yz} is performed
for the gauge coupling $\beta$. The fractional power
in eq. (\ref{reweight}) can be taken by using 
the fact that at $\mu=\mu_w$ the ratio of the determinants is 1 and
the ratio is a continuous function of the
chemical potential. The details of the determinant
calculation can be found in Ref. \cite{Fodor:2001pe}.

\TABLE{\centerline{\begin{tabular}{l|l|l|l|l|l}
$Re(\mu)$& 0.04 & 0.08 & 0.12 & 0.16 & 20\\
\hline
Re($\beta_0$); $L_s=6$ & 5.1863(9)
        &  5.1839(9)  &5.1800(9) &5.1749(11)       &5.1713(14)   \\
$10^2$Im($\beta_0$) &2.39(6)
        &2.39(6)  &2.41(8)       &2.41(13) &2.26(22)   \\
\hline
Re($\beta_0$); $L_s=8$ &5.1886(4)
        & 5.1858(5)        &5.1811(5)  &5.1753(7)       &5.1710(16)   \\
$10^2$Im($\beta_0$) &1.32(2)
        &1.33(3)  &1.33(3)  &1.28(5) &0.98(12)   \\
\hline
Re($\beta_0$); $L_s=10$ &5.1892(3)
        &5.1865(3)  &5.1821(3)  &5.1758(8) &5.1751(11)  \\
$10^3$Im($\beta_0$) &7.27(14)
        &7.26(15)   &7.33(22)   &6.44(74)  &5.29(76)   \\
\hline
Re($\beta_0$); $L_s=12$ &5.1888(2)
        &5.1861(2)   &5.1817(3)  &5.1768(4)  &5.1739(6)   \\
$10^3$Im($\beta_0$) &4.95(12)
        &4.95(13)   & 4.88(20) & 4.16(71)    & 2.07(74)  \\
\hline
Re($\beta_0$);$L_s\rightarrow \infty$ &5.1893(3)
        & 5.1866(3)        &5.1822(3) &5.1769(5) &5.1745(6) \\
$10^3$Im($\beta_0$) &2.12(14)
        &2.12(16)   &2.14(23)  &1.77(65)   &-0.39(77)  \\
\hline\hline
$\beta$ & $m_{u,d}$ & $m_\pi$        & $m_K$       & $m_\rho$ & $R_0$ \\
\hline
5.09 & 0.02        & 0.3555(1)       & 0.8948(2)   & 1.361(9) & 1.58(2)  \\
5.09 & 0.04        & 0.4978(1)       & 0.9235(1)   & 1.391(4) & 1.58(1)  \\
5.09 & 0.06        & 0.6044(1)       & 0.9511(1)   & 1.423(4) & 1.57(10) \\
\hline
5.16 & 0.02        & 0.3630(2)      & 0.9061(3)   &1.306(10)  & 1.73(3)  \\
5.16 & 0.04        & 0.5063(2)      & 0.9335(2)   &1.344(8)   & 1.67(1)  \\
5.16 & 0.06        & 0.6129(1)      & 0.9603(1)   &1.389(4)   & 1.64(1)  \\
\hline
5.19 & 0.02        & 0.3674(1)      & 0.9122(3)   & 1.287(8)  & 1.77(2)  \\
5.19 & 0.04        & 0.5063(1)      & 0.9337(2)   & 1.325(6)  & 1.72(1)  \\
5.19 & 0.06        & 0.6130(1)      & 0.9604(1)   & 1.366(2)  & 1.70(1)  \\
\hline
\end{tabular}}
\vspace{0.3cm}
\caption{\label{zeros}
$T\neq 0$ and $T=0$ results. The upper part is a
summary of the Lee-Yang zeros obtained at different chemical
potentials for $m_{u,d}$=0.0092 and $m_s$=0.25. 
We indicate by 6,8,10,12 and $\infty$ the spatial extensions --and their
extrapolation-- of our $L_t=4$ lattices. 
The lower part shows the
measured $T=0$ observables for three $\beta$ and three $m_{u,d}$ values
at $m_s$=0.25 on $12^3\cdot 24$ lattices.
}}

In the 
following we keep $\mu$ real and look for the zeros of the partition
function on the complex $\beta$ plane. These are the 
Lee-Yang zeros \cite{Yang:be}. Their V$\rightarrow \infty$
behavior tells the difference between a crossover and a first order
phase transition. At a first order phase transition the free energy
$\propto \log Z(\beta)$ is non-analytic. 
Clearly, a 
phase transition can appear only in the V$\rightarrow \infty$ limit, 
but not in a finite $V$. Nevertheless, the partition
function has Lee-Yang zeros at finite V. These are  
at ``unphysical'' complex values
of the parameters, in our case at complex $\beta$-s. For a 
system with a first order phase transition these zeros
approach the real axis in the V$\rightarrow \infty$ limit
(the detailed analysis suggests a $1/V$ scaling).   
This V$\rightarrow \infty$ limit generates the non-analyticity of
the free energy. For a system with crossover the free energy is analytic, 
thus the zeros do
not approach the real axis in the V$\rightarrow \infty$ limit.

\section{$T \neq 0$ and $T=0$ simulations for $n_f$=2+1.}

Using the formulation described above we study 2+1 flavour
QCD at $T\neq 0$ on $L_t=4,\ L_s=6,8,10,12$ lattices with
$m_{u,d}=0.0092$ and $m_s=0.25$ as
bare quark masses. Note, that these
mass parameters approximately correspond to their 
physical values. At $T=0$ we use $12^3\cdot 24$
lattices. Three different couplings ($\beta$=5.090, 5.160, 5.190)
are studied in order to determine the non-perturbative
$\beta$-function. Chiral extrapolation in the light quark
masses at $T$=0 are done by using three different mass 
parameters ($m_{u,d}$=0.02, 0.04, 0.06). 
For each parameter set 3000 configurations were generated.
For generating the
field configurations the R algorithm is applied. The microcanonical 
stepsize is always set to half of the light quark mass. We use a modified 
version \cite{Fodor:2002zi} of the
MILC collaboration's code \cite{milc}. 

At  $T\neq 0$
we determined the complex valued Lee-Yang zeros, 
$\beta_0$, for different V-s 
as a function of $\mu$. Their 
V$\rightarrow \infty$ limit was calculated by a 
$\beta_0(V)=\beta_0^\infty+\zeta/V$
extrapolation. The results (listed in Table \ref{zeros}) 
are obtained by generating  
100,000; 100,000; 100,000 and 150,000 configurations on our
$L_s$=6,8,10 and $12$ lattices, respectively. 
The determinant calculation was carried out after every 50 
trajectories. Thus our results are based on a few thousand independent
configurations.

Figure \ref{infV}
shows Im($\beta_0^\infty$) as a function of $\mu$ enlarged around 
the endpoint $\mu_{end}$.
The picture
is simple and reflects the physical expectations. For small
$\mu$-s the extrapolated Im($\beta_0^\infty$) is inconsistent with
a vanishing value, and the prediction is a crossover.
Increasing $\mu$ the value of Im($\beta_0^\infty$) decreases, 
thus the transition becomes consistent with a first order phase
transition. 
The statistical error
was determined by a jackknife analysis using subsamples of the
total $L_s$=6,8,10 and 12 partition functions. 
Our primary result in lattice units is $\mu_{end}=0.1825(75)$. 

\FIGURE{\centerline{\epsfig{file=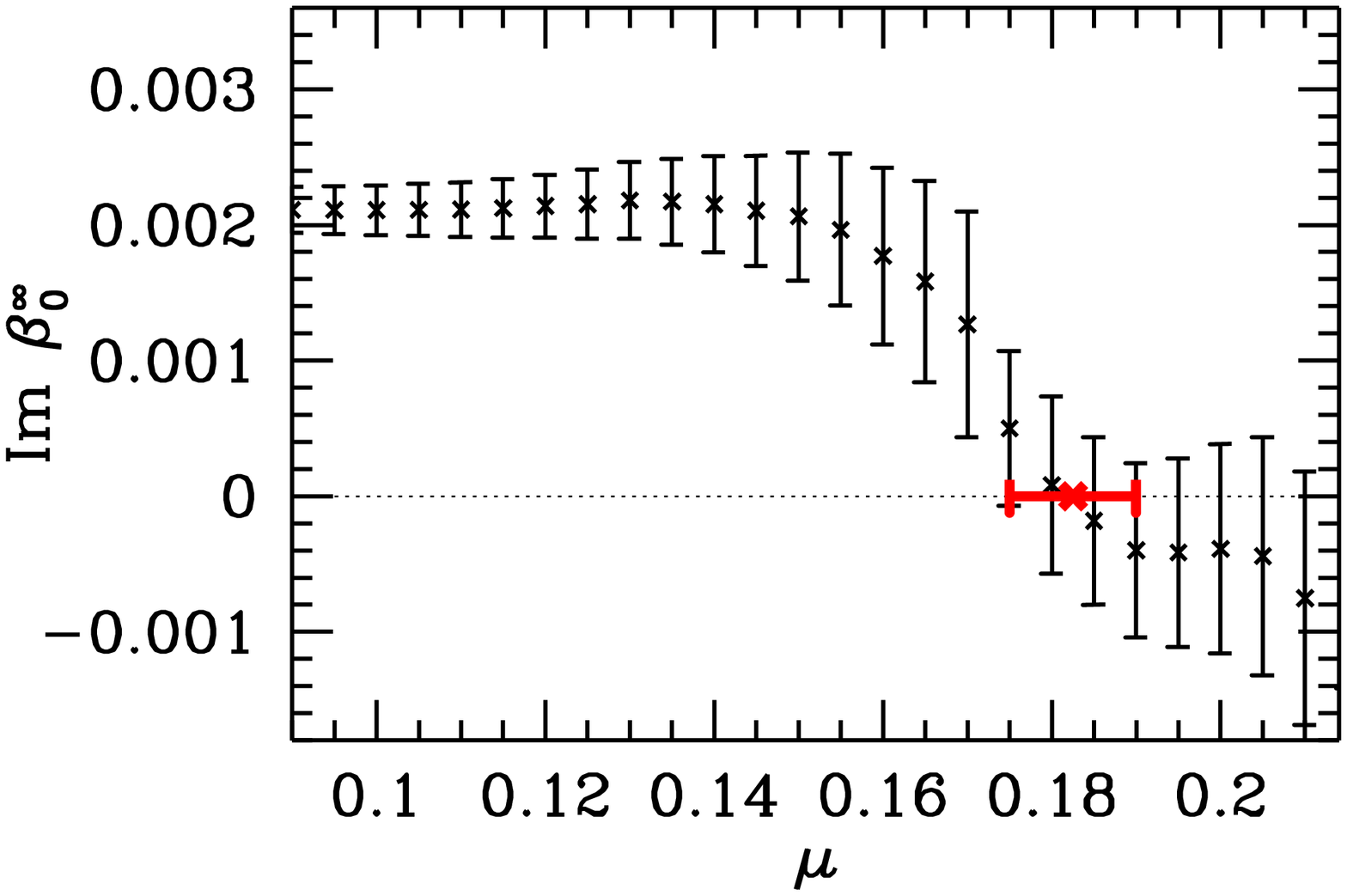,width=12cm,bbllx=0,bblly=265,bburx=580,bbury=630}}
\caption{\label{infV}
Im($\beta_0^\infty$) 
as a function of the chemical potential. 
}}
      
Table \ref{zeros} contains also the 
$T=0$ results. To set the physical scale we used an 
average obtained from $R_0$ (0.5~fm) and $m_\rho$ (770~MeV).
Note, that these two quantities give somewhat different scales
(the difference is fairly small, approximately 10-15\%). This difference
is expected to dissappear when one approaches the continuum limit
with physical quark masses.\footnote{
Note, that this difference is smaller than the difference in our previous 
analysis~\cite{Fodor:2001pe}. This improvement is due to the choice of 
the physical light quark masses.}

Setting the scale leads to the final results of the 
analysis. 
As we already discussed, the quark masses, used to 
determine the endpoint, correspond
approximately to their physical values. The pion to 
rho mass ratio, extrapolated to our $T\neq 0$ parameters, is 0.188(2) 
(its physical value is 0.179), whereas the
pion to K mass ratio in the same limit is 0.267(1) 
(its physical value is 0.277).   

Along the transition line $\beta$ decreases, thus the lattice
spacing increases. During the reweighting procedure 
we did not change the 
lattice quark masses, thus the quark masses changed in physical units.
As a consequence, the transition line slightly deviates 
from the line of constant physics. We corrected for this
effect by using our previous results at larger $m_{ud}$~\cite{Fodor:2001pe} 
and by following the observation that
transition lines for slightly different quark masses are
practically parallel (see \cite{Fodor:2002km,Csikor:2004ik}).

Let us estimate the applicability of the method approaching the
continuum limit.
In the present analysis with physical quark masses
the evaluation of the eigenvalues was
somewhat less costly than the production of the configurations.
Extending the analysis to even larger volumes, approximately upto
4$\cdot 16^3$, the eigenvalue determination remains subdominant. 
Using these larger volumes reduces the error on $\mu_{end}$ 
to a level, which is not even needed (uncertainties due to
finite lattice spacing are more important).
For finer lattices the eigenvalue evaluation 
goes with $L_s^9$ and the 
configuration production goes at least with $L_s^9$. According
to numerical estimates \cite{Fodor:2002km,Csikor:2004ik,Ejiri:2004yw}
the applicability range of the overlap-improving multi-parameter
reweighting technique, 
along the transition line,
scales with $\mu \propto V^\gamma$ 
with $\gamma\approx 1/3$. (Note, that the systematic study
of the lattice spacing dependence of $\gamma$ has not been performed 
yet.) In the scaling region the 
chemical potential of the endpoint
is constant in physical units. It scales with $V^{0.25}$ in
lattice units for fixed physical volumes. 
Taking into account the marginal difference between
$\gamma \approx 1/3$ and 0.25, one concludes that 
the determinant evaluation remains subdominant and even with the
present technique one might successfully approach smaller
lattice spacings. 

\section{Results and conclusions.}

Figure \ref{physical} shows the phase diagram 
in physical units, thus
$T$ as a function of $\mu_B$, the baryonic chemical potential 
(which is three times larger then the quark chemical potential). 

The transition temperature at vanishing baryonic chemical potential
is $T_c$=164$\pm$2~MeV. 
Note, that this value is somewhat 
smaller than our previous result ($T_c$=172$\pm$3~MeV of 
Ref. \cite{Fodor:2001pe}). 
This is a known phenomenon: smaller quark mass results in
smaller transition temperature
(see e.g. \cite{Karsch:2000ps}).

The curvature of the crossover line separating the QGP 
and the hadronic phases is given by $T/T_c=1-C\mu_B^2/T_c^2$ with
C=0.0032(1). This value is somewhat 
smaller than our previous
curvature \cite{Fodor:2001pe} or other values in the literature 
\cite{Allton:2002zi,Karsch:2003va,Allton:2003vx,deForcrand:2002ci}. 
Note, that compared to other analyses, 
we took into account an additional effect which reduced the value of $C$.
The small change of the mass 
parameter on the line of constant physics (caused by the change 
of the lattice spacing) slightly decreases the curvature.

The endpoint
is at $T_E= 162\pm 2$~MeV, $\mu_E=360 \pm 40$~MeV.
As expected, $\mu_E$ decreased as we decreased 
the light quark masses down to their physical values
(at approximately three-times larger $m_{u,d}$
the critical point was at $\mu_E$=720~MeV; see \cite{Fodor:2001pe}).

\FIGURE{\centerline{\epsfig{file=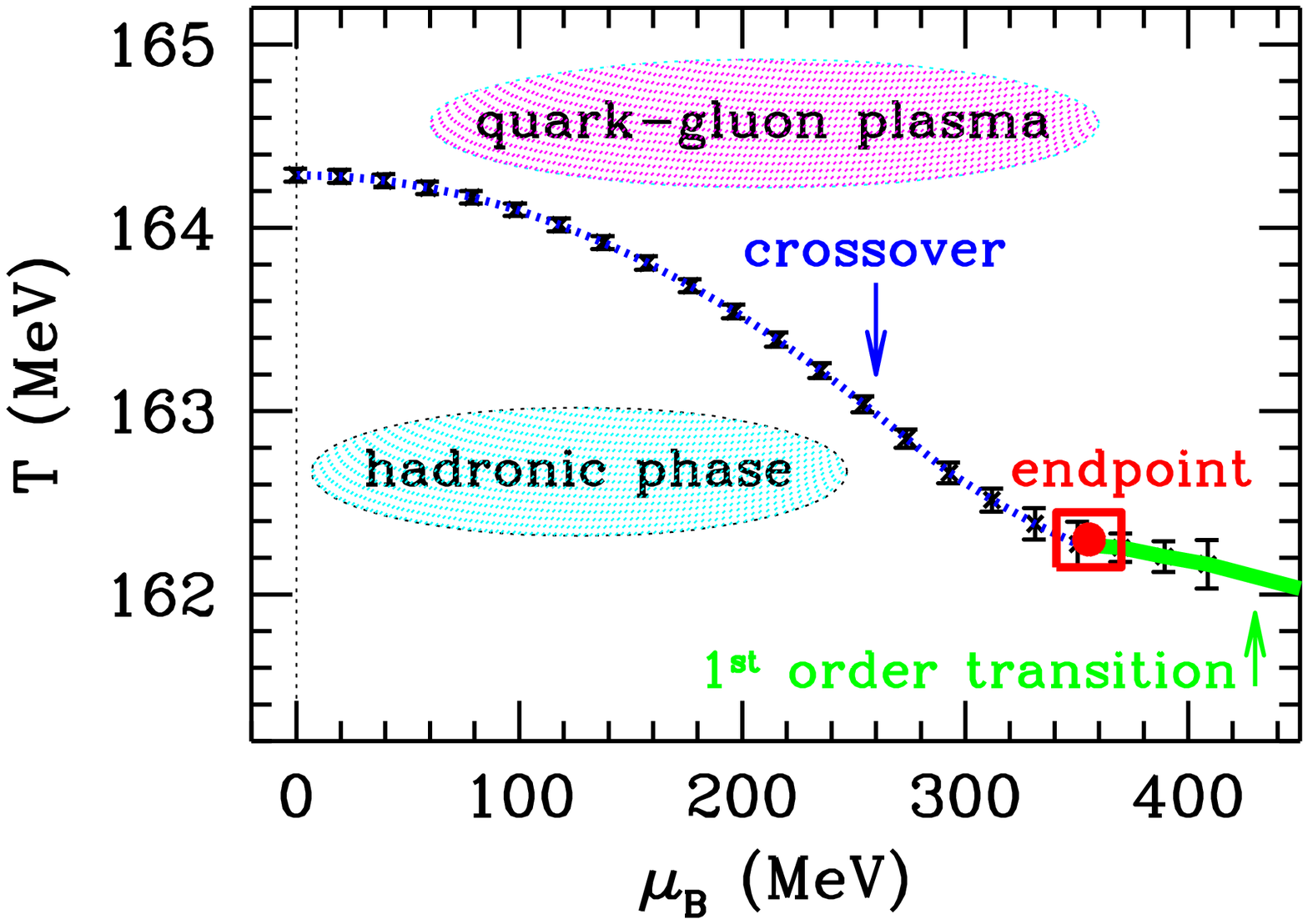,width=12cm,bbllx=0,bblly=265,bburx=580,bbury=630}}
\caption{\label{physical}
The phase diagram in physical units. 
Dotted line illustrates the crossover, solid line the first order phase
transition. The small square shows the endpoint.
The depicted errors originate from the reweighting procedure.
Note, that an overall additional error of 1.3\%
comes from the error of the scale determination at T=0. 
Combining the two sources of uncertainties one obtains
$T_E= 162\pm 2$~MeV and $\mu_E=360 \pm 40$~MeV.}}

The above result is a significant improvement
on our previous analysis \cite{Fodor:2001pe} 
by two means. We increased the physical volume
by a factor of three and decreased the light quark masses  
by a factor of three. 
Increasing the
volumes did not influence the results, which indicates
the reliability of the finite volume analysis. 
Clearly, more work is needed to get
the final values. Most importantly one has to extrapolate to 
the continuum limit.    

This work was 
partially supported by Hungarian Science Foundation grants No. 
OTKA-37615/\-34980/\-29803/\-M37071/\-OMFB1548/\-OMMU-708. This work was in 
part based on the MILC collaboration's lattice QCD code \cite{milc}.

\end{document}